\documentclass[sigconf]{acmart}
\AtBeginDocument{%
  \providecommand\BibTeX{{%
    \normalfont B\kern-0.5em{\scshape i\kern-0.25em b}\kern-0.8em\TeX}}}
\usepackage[english]{babel}

\usepackage{booktabs}
\usepackage{enumitem}
\usepackage{threeparttable}
\usepackage{float}
\usepackage{multirow}
\usepackage{textcomp}
\usepackage{caption}
\usepackage{booktabs}
\usepackage{bm}
\usepackage{algorithm}
\usepackage{algorithmic}
\usepackage{makecell}
\usepackage{url}
\usepackage[labelformat=simple]{subcaption}
\usepackage{amsmath}
\usepackage{graphicx}

\begin{document}
\title{DynamicRetriever: A Pre-training Model-based IR System with Neither Sparse nor Dense Index}
\thanks{* Equal contribution}
\thanks{$^\dagger$ Corresponding author}
\author{Yujia Zhou$^{1*}$, Jing Yao$^{1*}$, Zhicheng Dou$^{1\dagger}$, Ledell Wu$^{2}$, and Ji-Rong Wen$^{1}$}
\affiliation{%
  $^1$Gaoling School of Artificial Intelligence, Renmin University of China, Beijing, China \\
  $^2$Beijing Academy of Artificial Intelligence, Beijing, China
  \country{}
}
\email{{zhouyujia, jing_yao}@ruc.edu.cn}
\fancyhead{}

\begin{abstract}
Web search provides a promising way for people to obtain information and has been extensively studied. With the surgence of deep learning and large-scale pre-training techniques, various neural information retrieval models are proposed and they have demonstrated the power for improving search (especially, the ranking) quality. All these existing search methods follow a common paradigm, i.e. index-retrieve-rerank, where they first build an index of all documents based on document terms (i.e., sparse inverted index) or representation vectors (i.e., dense vector index), then retrieve and rerank retrieved documents based on similarity between the query and documents via ranking models. In this paper, we explore a new paradigm of information retrieval with neither sparse nor dense index but only a model. Specifically, we propose a pre-training model-based IR system called \textbf{DynamicRetriever}. As for this system, the training stage embeds the token-level and document-level information (especially, document identifiers) of the corpus into the model parameters, then the inference stage directly generates document identifiers for a given query. Compared with existing search methods, the model-based IR system has two advantages: i) it parameterizes the traditional static index with a pre-training model, which converts the document semantic mapping into a dynamic and updatable process; ii) with separate document identifiers, it captures both the term-level and document-level information for each document. Extensive experiments conducted on the public search benchmark MS MARCO verify the effectiveness and potential of our proposed new paradigm for information retrieval.
\end{abstract}

\maketitle
\section{Introduction}
Web search has become one of the main approaches for users to obtain information in their daily life. Given a query issued by the user, the information retrieval (IR) system retrieves a small number of relevant documents from massive web pages and returns a document ranking list as the final result. The traditional IR algorithms rely on the inverted index to complete the above process. The inverted index encodes term frequencies, term positions, document length, etc, of each document. Relevant documents are retrieved by counting the co-occurrence relationship between the query terms and document terms. 
The representative of this type of IR algorithm is the BM25~\cite{robertson2009BM25} model, which suffers from the challenge of word mismatching. With the development of natural language processing techniques, the understanding of terms has been elevated to the semantic level, alleviating the mismatch problem. Word embedding techniques, such as word2vec \cite{Mikolov2013Word2Vec}, allow models to measure the semantic similarity between any two words. Based on this, various neural matching models have been proposed to compute the relevance of query term sequence and document sequence~\cite{Xiong2017KNRM,Dai2018ConvKNRM}. They greatly improve search engine retrieval quality and user satisfaction.

Over the past few years, advances in representation learning leads to a shift from the traditional inverted index to dense vector index, where the IR system first encodes all documents into dense vectors and retrieves relevant documents based on the matching score between the vectors of the query and document. Recently, state-of-the-art pre-trained language models (PLM) show strong capability of involving contextual information to understand text sequences better~\cite{Devlin2019Bert,gpt,gpt-2,DBLP:conf/iclr/ClarkLLM20}. Motivated by this, some studies tried to explore the use of PLM for IR~\cite{DBLP:journals/corr/abs-1910-14424,DBLP:conf/iclr/ChangYCYK20,10.1145/3437963.3441777,DBLP:journals/corr/abs-1903-10972,DBLP:journals/corr/abs-2007-00808}, especially for the dense retrieval task. Considering that using matching tasks for pre-training is more suitable for the IR scenario, pseudo query-document pairs are constructed from the large corpus based on several strategies. These studies show that leveraging pre-trained language models can generate more accurate query and document representations to improve the retrieval performance~\cite{DBLP:journals/corr/abs-1910-14424,DBLP:journals/corr/abs-1901-04085,DBLP:conf/ecir/GaoDC21,DBLP:journals/corr/abs-1903-10972}.

Despite the great progress made by previous research, advanced IR models have the same framework, i.e. \textit{index-retrieval-rerank}, as traditional IR systems from decades ago, which includes three steps: (1) building an index for each document in the corpus; (2) retrieving a set of documents based on the query; (3) computing the relevance and re-ranking the candidate documents. This framework enables search engines to retrieve a small number of documents with low query latency, and then re-rank them through deep semantic matching. Recently, \citet{DBLP:journals/sigir/MetzlerTBN21} proposed that this fixed framework can be optimized using a unified model. Motivated by this blueprint, in this paper, we explore a new paradigm of information retrieval that eliminates both the sparse and dense index but only maintains a large-scale pre-training model. As for language models, the pre-training stage can be seen as the process of learning the basic meaning of each word and the dependencies between words. Similarly, for our model-based IR system, we can regard each document as an individual token and encode the knowledge of all documents in the corpus into the model through pre-training. With such a model being aware of both the semantic knowledge and document identifiers, we can complete a variety of downstream IR tasks, including document retrieval, response generation, document summarization, etc. Besides, this model enables all these tasks to be optimized end-to-end using a unified framework. In this paper, we mainly focus on exploring the task of document retrieval.

Specifically, we propose a pre-training model-based IR system with neither sparse not dense index, called DynamicRetriever. It is comprised of two modules: a PLM encoder to obtain the semantic representation of text sequences, i.e. queries and document passages, and a Docid decoder, which keeps a vocabulary of document identifiers and learns a vector for each docid. For a given query, DynamicRetriever first encodes this query into a context-aware representation vector, and then directly outputs document identifiers through the Docid decoder, which is different from previous index-based retrieval methods. This framework has several advantages compared with traditional index-based IR systems. \textbf{First}, the model-based approach parametrizes the traditional static index. This allows the model's understanding of the document content to be a dynamic process that can be updated during training. \textbf{Second}, unlike text-to-text matching, our model establishes a mapping from text to document identifier. Bridging the gap between terms and document identifiers can capture more document-level features such as authority and popularity, thereby enhancing the ranking quality. 


Similar to advanced language models, the training process of our model-based IR system includes pre-training and fine-tuning. At the pre-training stage, the semantic information of each document identifier can be memorized in the model through multiple pre-training tasks. At the fine-tuning stage, the model attempts to learn the \textit{query-docid} relations with labeled query-document pairs. Specifically, we implement two variants of the DynamicRetriever model. The first variant is called Vanilla model, which trains the Docid decoder module from scratch. However, since each document identifier is independent and learned separately, there is no obvious relatedness between the vector of two different document identifiers even they share similar semantic.
Due to the weak generalizability, the model struggles to predict documents correctly for those who lack fine-tuning data. Therefore, we propose the OverDense model, which combines the advantages of the model-based IR system and dense retrieval models. It uses existing vectorized indexes to initialize the Docid decoder, strengthening the model's understanding of each doc identifier.

We conduct experiments on MS MARCO to test the performance of the document retrieval task. Experimental results show that our proposed DynamicRetriever, which involves document identifiers into model parameters, is helpful for improving the retrieval results and has the potential to be scaled up.

In conclusion, the contributions of this paper are three-fold:
\begin{itemize}
    \item We explore a new paradigm of information retrieval that ditches building index but only maintains a large-scale pre-training model. Such model is promising to complete a variety of IR tasks and can be optimized end-to-end.
    \item Focusing on the document retrieval task, we propose a pre-training model-based IR system called DynamicRetriever. This model establishes the mapping from text to document identifiers directly to capture more document-level features.
    \item Under the model-based IR framework, we present two strategies to train the model with different initialization approaches, including using pre-training tasks or leveraging dense vectors to embeds the semantic of document identifiers.
\end{itemize}

\section{Related Work}
As stated in Section 1, existing information retrieval models follow the index-retrieve-rerank paradigm. The indexing and retrieval components are crucial and have been widely studied, especially in recent years when deep learning and large-scale pre-trained language models are developing. In this section, we briefly review the related works of this paper, including sparse retrieval and dense retrieval.

\subsection{Sparse Retrieval}
This is a traditional method for document indexing and retrieval. It first builds an inverted index based on all documents in the corpus, which encodes term frequencies, term position, document structure information, document length and so on. Then, it retrieves relevant documents based on the matching between query terms and document terms. How to measure the relevance between terms and the weights of different terms is the main challenge for sparse retrieval. The classical TF.TDF and BM25~\cite{robertson2009BM25} methods employ the term frequency and precise word matching, achieving great results. Furthermore, several works apply neural networks to improve the performance of sparse retrieval from the semantic aspect. Word embedding techniques such as Word2Vec~\cite{Mikolov2013Word2Vec} are introduced to better measure the semantic similarity between different query terms and document terms~\cite{nogueira2019SparseVector}, alleviating the mismatch problem. Some deep learning based models\cite{Nogueira2019SparseExpansion,Raffel2020Transformer} also expand possible terms for the issued query to improve the recall. DeepCT~\cite{Dai2019DeepCT} employs the large-scale language model BERT~\cite{Devlin2019Bert} to predict the term weights, instead of traditional term frequency.

\begin{figure*}
    \centering
    \includegraphics[width=0.9\linewidth]{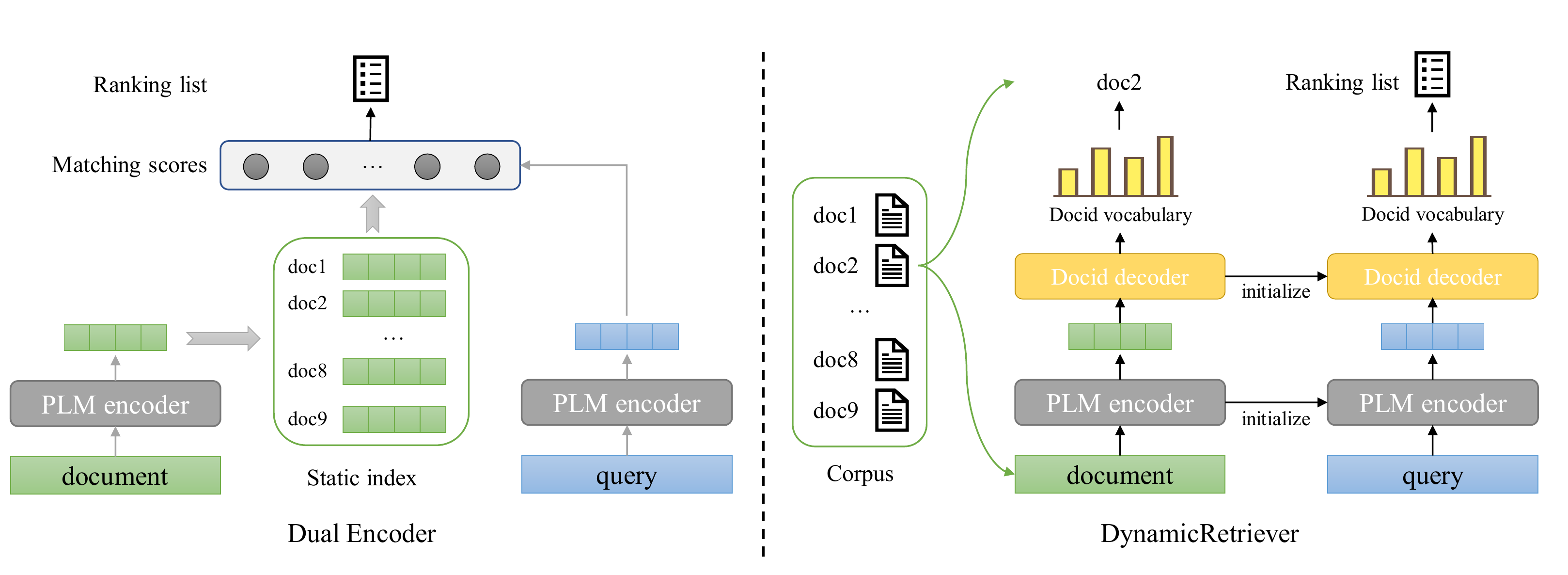}
    \caption{Comparison between dual encoder model and DynamicRetriever}
    \label{fig:basemodel}
\end{figure*}
\subsection{Dense Retrieval}
Dense retrieval is a representation-based method for indexing and retrieval. First, it applies a neural network to encode each of the documents into a dense vector and builds a vectorized index. Then, it embeds the issued query into the same latent space and computes the similarity between the query representation and document vectors to efficiently retrieve relevant documents~\cite{Johnson2021SimSearch}, where inner product, cosine similarity and efficient K-nearest neighbor search could be employed. Compared with the sparse retrieval, encoding the query and documents into low-dimensional vectors for matching is promising to capture rich semantic and contextual information and provide a way to alleviate the vocabulary mismatch problem. However, due to the precise match between tokens is ignored, the precision may be sacrificed. With the development of neural models and large-scale pre-trained language models to better learn contextual information of documents, such as Transformer~\cite{Vaswani2017Transformer}, Bert~\cite{Devlin2019Bert} and GPT~\cite{gpt}, dense retrieval is receiving more and more attention, which is demonstrated to outperform sparse retrieval~\cite{Guo2020Dense,Lee2019Dense,Zhan2020Dense,Chang2020Dense}. Some studies attempt to explore the use of PTM for IR~\cite{DBLP:journals/corr/abs-1910-14424,DBLP:journals/corr/abs-1901-04085,DBLP:conf/ecir/GaoDC21,DBLP:journals/corr/abs-1903-10972}. They design various strategies to construct pseudo query-document pairs from the large corpus and pre-train a model to generate query and document representations more accurately, improving retrieval performance.

Different from the above retrieval approaches with a separate index, we explore a new search paradigm and propose a model-based IR system with neither sparse nor dense index in this paper.

\section{DynamicRetriever: a pre-training model-based IR framework}
Existing search methods follow an \textit{index-retrieval-rerank} framework that has dominated IR systems for decades. They first attempt to encode the document content into sparse inverted index or dense vector index, then retrieve and rerank retrieved documents based on similarity between the query and documents, where index is always a necessary part. With the advent of pre-training techniques, we envision that the model can involve both the semantic information and corresponding document identifiers through pre-training, thereby replacing traditional static indexes. In such a model-based IR system, given a query, the document identifiers can be directly generated as the result. In this section, we propose a basic pre-training model-based IR system named DynamicRetriever, and implement two variant models. The details are introduced in the next.

\subsection{Model Architecture}
The whole architecture of DynamicRetriever is shown in the right part of Figure~\ref{fig:basemodel}. It works in two main steps: given a query, the model first encodes it into a vector with a PLM encoder, and then maps it to the docid vocabulary through a Docid decoder, outputting document identifiers. Compared with the dual encoder model presented in the left part of Figure~\ref{fig:basemodel}, which computes static vectorized index for all documents using a trained encoder, DynamicRetriever stores the semantic information and all document identifiers in the parameters of the Docid decoder. This can be viewed as the dynamic index that is updated directly as the model training. In the following, we describe the workflow of DynamicRetriever in detail.

First, given a query containing $n$ tokens, i.e. $q=\{w_1,w2_,...w_n\}$, we are supposed to understand this query for analyzing user's information need. The queries issued by users are often very short, which leads to a lot of ambiguity for understanding. Therefore, it is crucial to model queries in a fine-grained way. We apply a Transformer-based PLM encoder to compute the sentence embedding of $q$, denoted as: 
\begin{equation}
    V^q=\text{Transformer}^{cls}([w_1,w_2,...w_n]).
\end{equation}
We take the output of \textit{cls} as the query representation $V^q$.

Second, with the encoded query representation, the target of our model is to directly generate the most relevant document identifiers in the entire corpus. To implement this goal, we feed $V^q$ into the Docid decoder to obtain a probability distribution over all the document identifiers. Formally, assuming that there are $|D|$ document identifiers in the corpus, the probability distribution is calculated simulating the output layer of generative language models:
\begin{equation}
    O^q=\text{softmax}(W_{doc}^\mathrm{T} \cdot V^q),
\end{equation}
where $W_{doc} \in \mathbb{R}^{d_{model} \times |D|}$ is an project matrix to map the query representations to the probability of each docid. It can be viewed as dynamic indexes that can be updated during the model training. According to the output $O^q$, we are able to retrieve the top-k document identifiers by sorting the probability for the given query $q$. 

\begin{figure*}
    \centering
    \includegraphics[width=0.9\linewidth]{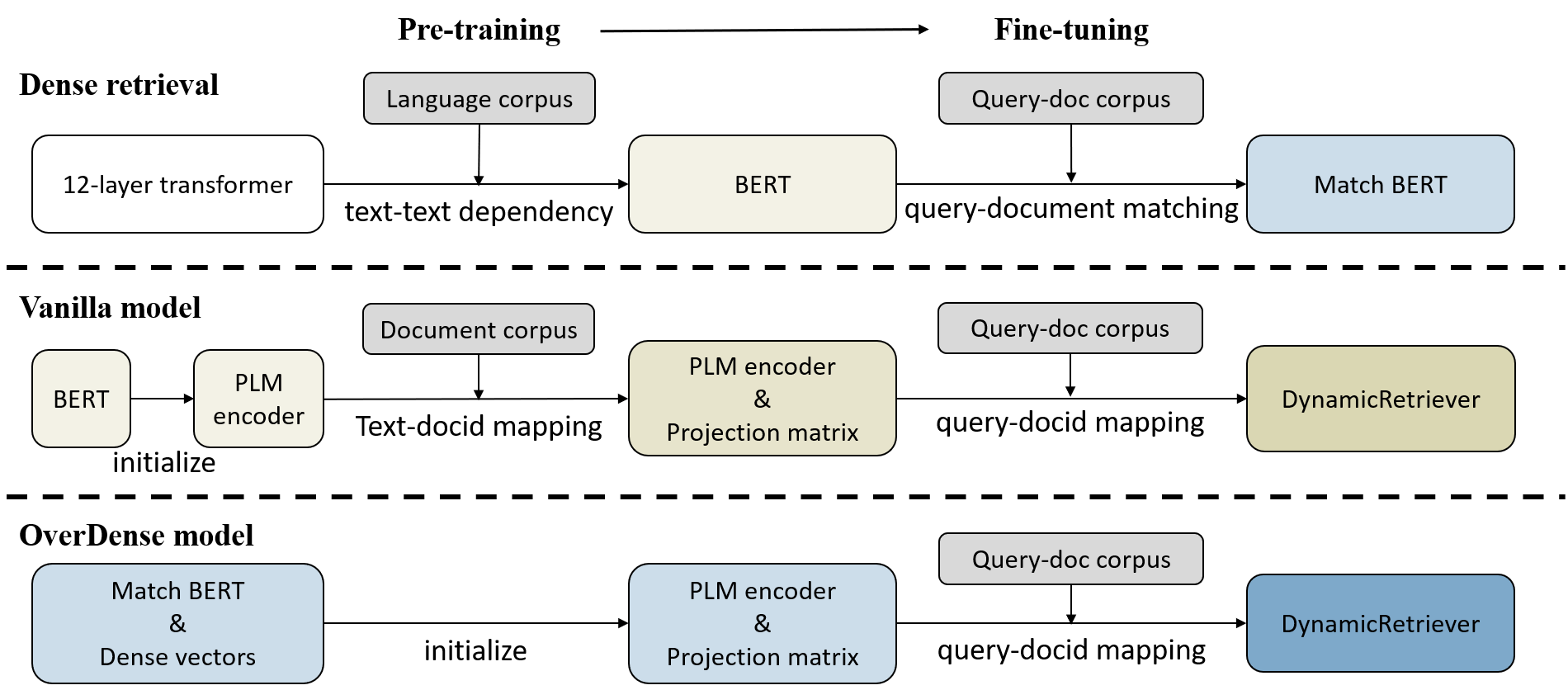}
    \caption{Model training workflows of dense retrieval and two variant models of DynamicRetriever.}
    \label{fig:framework}
\end{figure*}

\subsection{Encoding Document Identifiers into Model}
The training of pre-trained language models such as BERT concludes the pre-training stage and the fine-tuning stage. The pre-training of the language model focuses on learning the basic semantics of words and the semantic dependencies between words. At the fine-tuning stage, the model will enhance the ability to handle specific tasks. Similar to these PLMs, at the pre-training stage of DynamicRetriever, we hope the semantic information of each document identifier can be memorized in the model through multiple pre-training tasks. The fine-tuning stage is used to learn the matching relationships between queries and document identifiers. During this process, a large number of document-level meta information can be captured over term-level semantics, thereby improving the ranking quality. Under such a training framework, we propose two variant models with different training strategies: (1) Vanilla model, which warms up the model parameters with pre-training tasks and fine-tunes the model over the query-docid matching data. (2) OverDense model, which initializes the Docid decoder parameters over trained dense vectors and then continues training with query-docid relations. The workflows of current dense retrieval method and the two variant models of DynamicRetriever are shown in Figure~\ref{fig:framework}. 

\subsection{Vanilla Model: Training from Scratch}
The vanilla model initializes the projection matrix $W_{doc}$ randomly and trains it from scratch. Firstly, we devise three pre-training tasks to encode the semantic of each document identifier into the model. Then, we use labeled query-document pairs to fine-tune the model parameters by capturing document-level features. 

\begin{figure}
    \centering
    \includegraphics[width=0.9\linewidth]{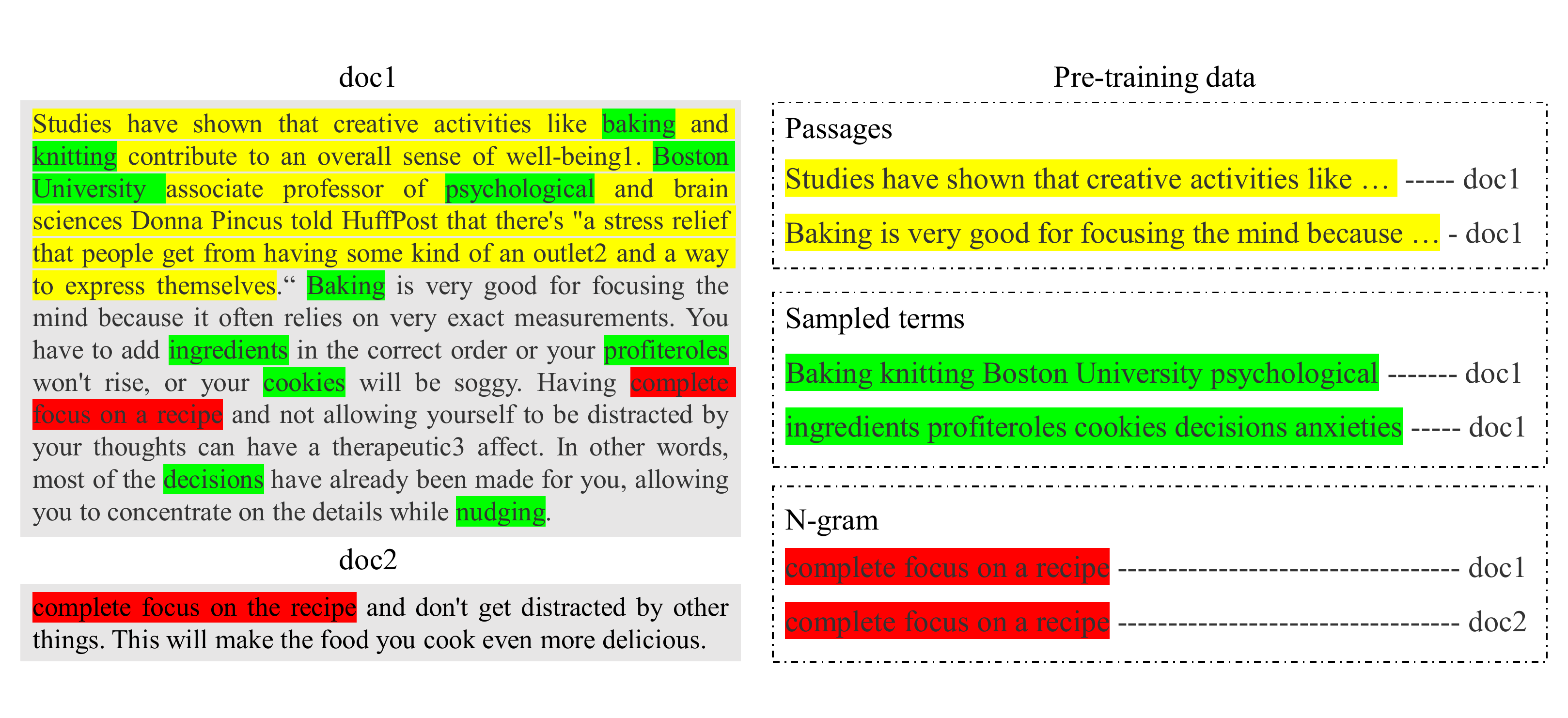}
    \caption{Pre-training tasks of the Vanilla model.}
    \label{fig:pretrain}
\end{figure}
\textbf{Pre-training}. This stage is designed to learn the knowledge from the large corpus, which is used to pre-train the semantic of each docid in the large corpus. A critical step is to extract self-supervised signals from the corpus and construct (term sequence-docid) pairs. We try three strategies that may contribute to pre-training. As shown in Figure~\ref{fig:pretrain}, they are:
\begin{itemize}[leftmargin=*]
    \item \textbf{Training with passages}. Previous studies have shown that using passage-level evidence for document ranking can enhance the ranking quality~\cite{DBLP:conf/sigir/Callan94}. Inspired by this, we attempt to segment the document text into multiple passages with fix-sized windows. Each passage can reflect a local view of the document content. For the document \textit{doc1}, whose content can be divided into $m$ passages, we can construct $m$ pairs for training, i.e. ($\text{passage}_1,doc1$),...,($\text{passage}_m,doc1$).
    
    \item \textbf{Training with sampled terms}. The importance of each word in the document is different, and often some important words can reflect the basic content of the document. Therefore, we sample terms according to the word importance with random length (from 10 to 512). Formally, for the document \textit{doc1}, after $m$ times of sampling, we obtain $m$ sets of terms for training, i.e. ($\text{set}_1,doc1$),...,($\text{set}_m,doc1$).
    
    \item \textbf{Training with ngram}. Ngram is a sequence of $n$ words, which may appear in multiple documents. To some extent, this sequence can characterize the similarity between multiple documents. It can be seen as an enhanced version of the inverted table. For a sequence of $n$ words, supposing we find it in $m$ documents, then the training pairs can be formed as ($\text{n-gram},doc1$),...,($\text{n-gram},docm$).
    
\end{itemize}

\textbf{Fine-tuning}. After pre-training, the model has memorized the basic semantic of each docid. The fine-tuning stage is used to learn the query-docid relations with supervised matching data. Different from traditional two-tower matching models focusing on text matching, our model pays more attention to bridging the gap between terms and document identifiers through training. For the query $q$, it's representation is denoted as $V_q$. We choose cross entropy as the loss function: 
\begin{equation}
    \mathcal{L}=\sum_i y_i\cdot \frac{\text{exp}(u_i^\mathrm{T} \cdot V^q)}{\sum_{j=1}^{|D|} \text{exp}(u_j^\mathrm{T} \cdot V^q)},
\end{equation}
where $u_i$ is the $i_{th}$ column of projection matrix $W_{doc}$.

However, there are two shortcomings of the vanilla model. (1) The learning of mapping relations from query text to docid is overly dependent on the fine-tuning task. With limited fine-tuning data, a large number of document identifiers can only be learned from a small number of samples in the pre-training tasks. If we lack enough fine-tuning data, the model will perform poorly. (2) The dependencies between docids are difficult to capture, which leads to poor generalization ability of the model. To overcome these issues, we propose the OverDense model to incorporate the benefits of dense retrieval into the model-based IR system.

\subsection{OverDense Model: Training over Dense Vectors}
By comparing DynamicRetriever and dense retrieval models, we find that their advantages are complementary. For example, a major advantage of dense retrieval is that it can enhance the generalization of the model. Moreover, it is good at extracting term-level features in a fine-grained manner. These advantages are exactly what our model lacks. If we combine the advantages of them, a model with strong generalizability and multi-level feature extraction can be trained to enhance the performance. Based on this consideration, we attempt to integrate the advantages of dense retrieval models into our framework. We devise the OverDense model which has different parameter initialization strategies compared with the vanilla model.

To initialize the parameters of each document identifier in the model, the Vanilla model constructs self-supervised data to learn the text-docid relations. However, this approach is difficult to exploit the relationship between document identifiers. Therefore, we propose a new framework to train the model, which has three steps:

\begin{itemize}
    \item \textbf{Fine-tuning the two-tower BERT with query-document pairs}. Two-tower BERT is a typical framework of PLM-based dense retrieval. Endowed with the benefit of PLM's powerful semantic modeling capability, this framework greatly improves retrieval quality while enhancing the generalization ability of the model compared to sparse retrieval. To strengthen the performance of BERT on the query-document matching task, we use labeled query-document pairs to fine-tune the two-tower BERT, which will be used in the next.
    \item \textbf{Generating dense vectors to initialize the model parameters}. After fine-tuning the BERT for matching, we can compute the dense vectors of each document. These vectors fully integrate the textual semantic information of the documents, so documents with similar texts have higher vector similarity. If we initialize the projection matrix of the Docid decoder with dense vectors, the problem of poor model generalization can be alleviated. After initializing $W_{doc}$ with dense vectors, our model can achieve the same performance as dense retrieval. Continuing to train on this basis, our model can pay more attention to document-level information, thereby improving the model performance on document retrieval.
    \item \textbf{Fine-tuning our model with query-docid pairs}. The initialized parameters have fully recorded the semantic information of each docid, and also have integrated the text-text matching information. Next, we will extract text-docid matching relations from supervised data, and continue training the model. We expect the model can capture document-level features such as authority. For example, there are a lot of forwarded news on the Internet, the content is roughly the same but the publishers are different. Users tend to choose the more authoritative one. Traditional PLM-based methods try to model such information at the term-level, which is an indirect way with information loss. In contrast, our model can model the document-level features by directly updating the representation of document identifier without relying on terms.
\end{itemize}

\subsection{Discussion}~\label{subsec:discussion}
The model-based IR system uses model parameters in place of traditional static document indexes. A natural problem is how to scale the model to a larger corpora. As the number of documents increases, the model has to use more and more parameters to memorize document identifiers. Due to memory constraints, the number of parameters of our single model cannot grow infinitely. This prompts us to think about how to deal with large-scale corpus scenarios. There are two potential solutions to this problem: distributed model and hierarchical model. 

\begin{figure}
    \centering
    \includegraphics[width=0.8\linewidth]{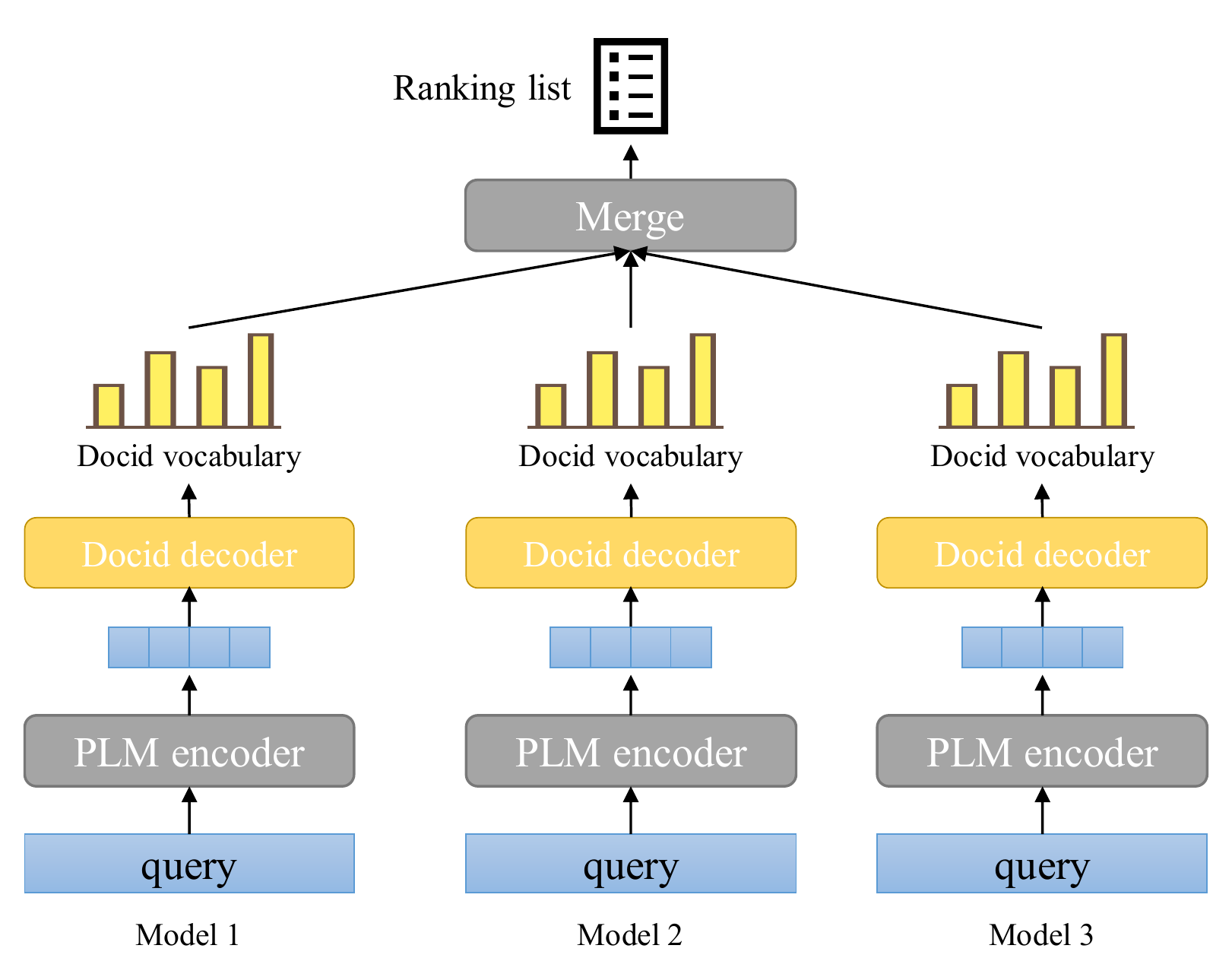}
    \caption{The architecture of the distributed model.}
    \label{fig:distributed}
\end{figure}
\textbf{Distributed model}. Now that our model can work on small scale data, we can train multiple sub-models distributedly, and then fuse their predictions to get the final document ranking list. As shown in Figure~\ref{fig:distributed}, we devise multiple models with the same structure, and each model is responsible for learning the mapping relationships of a part of the document identifiers. Finally, given a query, each model can compute a probability distribution over some document identifiers. By a merge function, we can get the whole probabilities and generate the most relevant docid. However, merging the outputs in such a simple way may cause another problem. Due to different sub-models are trained independently, the scale of document scores of different sub-models are not consistent. Therefore, more suitable merge functions or training strategies are needed. A possible solution is that we can add some common documents into different sub-models, so as to scale the space of each one to the same level. This question requires more exploration in the future.

\textbf{Hierarchical model}. Another line to solve this problem is to reduce the number of document identifiers by using multi-segment identifiers to represent a document. For example, if we use two parts to represent a document, where the first part represents the category and the second part represents the id under that category, we just need $log(|D|)$ ids to cover all documents. So we organize and categorize documents in a structured way, and then encode them into strings as docids by category. To classify documents, the easiest way is to randomly group them. But if we want each category id to have a specific meaning, more sophisticated approaches are needed, such as grouping documents based on domains of url, or clustering semantic representations of all documents. We believe that different classifications will affect the training process of the model. A reasonable and effective classification method deserves to be further explored.

\section{Experimental Settings}
\subsection{Dataset}
MS MARCO (Microsoft Machine Reading Comprehension) is a large-scale dataset collected from the field of machine reading comprehension, which is widely used on various tasks after its release, including question answering, passage ranking, document ranking and so on. In this paper, our experiments about the document ranking task are conducted on the MS MARCO dataset and we mainly focus on the results of ad-hoc retrieval. This dataset contains a total of 3.2 million candidate documents with a mean document length of 1,600 terms. The training set has 367,013 queries, while the testing set has 5,193 queries. For each training or testing query, there is a positive document in the document set, which contains at least one passage manually annotated as positive to answer the corresponding query. All the testing queries are distinguished from the training queries, and there is little overlap between their corresponding positive document sets. According to the above statistics, about 300K+ documents have been clicked and can be used to construct query-document pairs for model training. Thus, to evaluate the performance and scalability of our model on datasets with different distributions, we consider different sampling methods to construct candidate document sets and conduct extensive experiments for comparison. At first, we rank all candidate documents based on their click frequency and select the top 100K, 200K, 300K documents to construct different document sets for evaluation (Corresponding to Top 100K, 200K, 300K in Table~\ref{tab:dataset}). In addition, we also randomly sample 100K, 200K, 300K documents from the whole document set for testing (Corresponding to Rand 100K, 200K, 300K in Table~\ref{tab:dataset}), as well as the whole set with 3.2 million documents. As for these different documents subsets, we only consider the training queries and testing queries whose clicked documents exist in this set. Statistics of different data subsets are presented in Table~\ref{tab:dataset}.
\begin{table}[]
    \centering
    \caption{Statistics of MS MARCO and different data subsets.}
    \begin{tabular}{c|ccrr}
    \toprule
         Dataset & \#Doc & \#Passage & Train Pairs & Valid Pairs\\
    \midrule
         Top 100K & 100K & 955,586 & 147,086 & 466 \\
         Top 200K & 200K & 1,763,726 & 247,086 & 636 \\
         Top 300K & 300K & 2,721,974 & 347,086 & 778 \\
         Rand 100K & 100K & 838,527 & 11,262 & 156 \\
         Rand 200K & 200K & 1,656,273 & 22,907 & 317 \\
         Rand 300K & 300K & 2,477,582 & 34,290 & 487 \\
         MS MARCO & 3.2M & 25,600,715 & 367,013 & 5,193 \\
    \bottomrule
    \end{tabular}
    \label{tab:dataset}
\end{table}

\subsection{Evaluation Metrics}
Due to that we focus on the results of the document retrieval task, we use the metric Recall@k where k=\{1,20,100\} to evaluate the recall power of our model and all the baselines.
In addition, we also pay attention to the document ranking quality and apply MRR for evaluation.

\subsection{Baselines}
In order to confirm the effectiveness of our model, we select several baselines for comparison, including the classical BM25 algorithm for sparse retrieval and the recent dense retrieval methods based on BERT.

\textbf{BM25}: It~\cite{robertson2009BM25} is a bag-of-word retrieval method that ranks the candidate documents based on the TF-IDF weights of the query terms appearing in each document, traditional but effective. 

\textbf{BERT}: With the development of large-scale pre-trained language models such as BERT, a two-tower framework for dense retrieval becomes popular, which encodes the query and document into a representation vector respectively and computes the dot product between them as the ranking score. We first use the training query-document pairs to fine-tune the parameters of BERT and generate the representation vectors for all candidate documents. When testing, we encode each query into a vector and retrieve the documents with the largest ranking scores. For all texts inputted into BERT, we keep the first 512 tokens~\cite{Devlin2019Bert,Zhan2020Dense}.

\textbf{D-Vanilla} and \textbf{D-OverDense} indicate the two variants of our proposed DynamicRetriever IR system respectively.

\subsection{Implementation Details}
In our DynamicRetriever model, the query understanding component is initialized by the pre-trained bert-base-uncased model (12 layers, hidden states 768d) from the Transformers\footnote{https://huggingface.co/bert-base-uncased/tree/main}, and the document prediction module is a $768\times N$ linear layer where $N$ corresponds to the size of the candidate document set. The max length of tokens inputted into the BERT encoder is set as 512. The length of sampled terms could be from 10 to 512. For both the pre-training and fine-tuning stages, we apply AdamW to optimize the parameters with the learning rate as 5e-5. All experiments are completed with NVIDIA-V100(32 GB). The batchsize is set differently when different sizes of document set are considered.

\section{Experimental Results}
We conduct extensive experiments to confirm the advantages of our proposed DynamicRetriever. In this section, we present the experimental results and make some analyses.

\begin{table*}[t!]
    \centering
    \caption{Overall performance of all baselines and our proposed models. ``$\dagger$'' denotes the result is significantly better than other models from the same setting in t-test with $p \textless 0.05$ level. The best results are in \textbf{bold}.}
    \begin{tabular}{p{0.1\linewidth}|p{0.05\linewidth}p{0.06\linewidth}|p{0.05\linewidth}p{0.06\linewidth}|p{0.05\linewidth}p{0.06\linewidth}|p{0.05\linewidth}p{0.06\linewidth}|p{0.05\linewidth}p{0.06\linewidth}|p{0.05\linewidth}p{0.06\linewidth}}
    \toprule
     \multirow{2}[2]{*}{Model} &  \multicolumn{4}{c|}{Top 100K} & \multicolumn{4}{c|}{Top 200K} & \multicolumn{4}{c}{Top 300K} \\
        \cmidrule(lr){2-13}
        & \multicolumn{2}{c|}{Recall@20} & \multicolumn{2}{c|}{MRR} & \multicolumn{2}{c|}{Recall@20} & \multicolumn{2}{c|}{MRR} & \multicolumn{2}{c|}{Recall@20} & \multicolumn{2}{c}{MRR}\\
        \midrule
        BM25 & 0.5483 & -33.82\% & 0.2811 & -33.67\% & 0.4685 & -41.65\% & 0.1968 & -51.01\% & 0.4185 &	-49.02\% & 0.1743 & -58.21\% \\
        BERT & 0.8281 & - & 0.4238 & - & 0.8029 & - & 0.4017 & - & 0.8203 & - & 0.4171 & - \\
        D-Vanilla & 0.8784$^\dagger$ & 6.04\% & 0.5637$^\dagger$ & 33.01\% & 0.7562 & -5.74\% & 0.4616 & 14.91\% & - & - & - &  \\
        D-OverDense & \textbf{0.8861}$^\dagger$ & 7.00\% &  \textbf{0.5728}$^\dagger$ & 35.16\% & \textbf{0.8706}$^\dagger$ & 8.48\% &  \textbf{0.5221}$^\dagger$ & 29.97\% & \textbf{0.8525}$^\dagger$ & 3.90\% & \textbf{0.4877}$^\dagger$ & 16.93\% \\
     
    \midrule
    \midrule
     \multirow{2}[2]{*}{Model} &  \multicolumn{4}{c|}{Rand 100K} & \multicolumn{4}{c|}{Rand 200K} & \multicolumn{4}{c}{Rand 300K} \\
        \cmidrule(lr){2-13}
        & \multicolumn{2}{c|}{Recall@20} & \multicolumn{2}{c|}{MRR} & \multicolumn{2}{c|}{Recall@20} & \multicolumn{2}{c|}{MRR} & \multicolumn{2}{c|}{Recall@20} & \multicolumn{2}{c}{MRR}\\
        \midrule
        BM25 & 0.5823 & -26.33\% & 0.3606 & -33.99\% & 0.5201 & -25.29\% & 0.3106 & -29.49\% & 0.4864 & -23.82\% & 0.2811 & -24.48\% \\
        BERT & 0.7901 & - & 0.5463 & - & 0.6965 & - & 0.4405 & - & 0.6389 & - & 0.3722 & - \\
        D-Vanilla & 0.6922 & -12.41\% & 0.4985 & -8.75\% & 0.2126 & -69.54\% & 0.1432 & -67.49\% & 0.1225 & -80.88\% & 0.1036 & -72.17\% \\
        D-OverDense & \textbf{0.8423}$^\dagger$ & 6.58\% &  \textbf{0.6445}$^\dagger$ & 17.98\% & \textbf{0.7825}$^\dagger$ & 12.36\% & \textbf{0.4953}$^\dagger$ & 12.44\% & \textbf{0.6988}$^\dagger$ & 9.40\% & \textbf{0.4085}$^\dagger$ & 9.75\% \\
    \bottomrule
    \end{tabular}
    \label{tab:overall}
\end{table*}
\subsection{Overall Performance}
To start with, we compare DynamicRetriever to the selected baselines on various data subsets with different scales and data distributions to verify its effectiveness and scalability. The results are illustrated in Table~\ref{tab:overall}. We come to several conclusions as follows.

(1)	On all data subsets, our proposed DynamicRetriever achieves better results than BM25 and the BERT based two tower model. The D-OverDense model performs the best, which significantly outperforms all baselines with paired t-test at p<0.05 level. BM25 applies traditional sparse index and retrieves relevant documents based on precise matching between the query terms and document terms, while the BERT-based two tower model is state-of-the-art for dense retrieval which builds a vectorized index based on the document tokens. In our DynamicRetriever, the document index is parameterized and embedded into a large-scale model, implementing a model-based IR system. We analyze the reasons why our new method achieves better results exactly correspond to its advantages: 1) DynamicRetriever captures both the token-level and document-level information such as authority and popularity for each document; 2) it keeps a dynamic index of all documents that can be updated during model training.

(2) Observing the results of the D-Vanilla model, we find that this intuitive strategy can achieve great performance on small subsets. Specifically, on the Top 100K subset, the D-Vanilla model improves BERT by 6.04\% on the evaluation metric Recall@20, and 33.01\% on MRR. However, as the data scale increases, the performance of the D-Vanilla model drops sharply, especially for the subsets with rand 200K, 300K documents. We analyze the possible reason is that different from the previous dense retrieval framework using token-level information, our document-level DynamicRetriever system regards each document independently, thus the corresponding training data for each document is much less than that for each token. With the increase of document corpus size, the difficulty of distinguishing between documents increases, so the retrieval effect naturally decreases.

(3) Compared with the D-Vanilla model, the performance of our improved D-OverDense model is much better. Whether on the data subset of 100K documents or the later expanded 200K and 300K subsets, the performance of the OverDense model is consistently better than all baselines, showing strong scalability. Especially on subsets with 200K and 300K documents, the D-OverDense's results are significantly improved compared to the D-Vanilla model. Unlike the D-Vanilla model, which uses the pre-training tasks we design to learn the parameters for indexing documents, the OverDense model uses fine-tuned two-tower BERT model to generate the document representations for initializing this part of parameters. Then, a further fine-tuning task based on Q-D pairs is conducted to make fuller use of token-level and document-level information. Therefore, the performance of D-OverDense can be greatly improved.

In summary, the experimental results prove that \textbf{our proposed DynamicRetriever, which considers using model parameters as dynamic document indexes and capturing document-level information, is helpful for improving document retrieval results}.

\begin{table}[t!]
    \centering
    \caption{Ablation study of our models. `w/o fine-tune' is to test the model performance on zero-shot learning.}
    \begin{tabular}{p{0.3\linewidth}|p{0.12\linewidth}p{0.12\linewidth}|p{0.12\linewidth}p{0.12\linewidth}}
    \toprule
     \multirow{2}[2]{*}{Model} &  \multicolumn{4}{c}{Top 100K} \\
        \cmidrule(lr){2-5}
        & \multicolumn{2}{c|}{Recall@20} & \multicolumn{2}{c}{MRR}\\
        \midrule
        D-Vanilla & 0.8784 & - & 0.5637 & -\\
        \;\;w/o Pre-train & 0.0100 & -98.86\% & 0.0019 & -99.66\% \\
        \;\;w/o Fine-tune & 0.5323 & -39.41\% & 0.2901 & -48.54\%\\
        \midrule
        D-OverDense & 0.8861 & - & 0.5728 & -\\
        \;\;w/o Fine-tune & 0.8281 & -6.55\% & 0.4238 & -26.01\%\\
    \bottomrule
    \end{tabular}
    \label{tab:ablation_study}
\end{table}

\subsection{Ablation Study}
In our DynamicRetriever, there are several pre-training tasks and a fine-tuning task to parameterize the index of documents. We conduct ablation studies to analyze the effect of each task and display the results in Table~\ref{tab:ablation_study}.

We consider dropping several tasks listed as follows.

\textbf{D-Vanilla w/o Pre-train}: as for the D-Vanilla model, we train it with passages segmented from each document, sampled terms and n-grams to embed to document information into model parameters. We discard this pre-training task for verification.


\textbf{D-Vanilla w/o Fine-tune}: We skip the fine-tuning stage with Q-D Pairs and directly evaluate the D-Vanilla model after the pre-training.

\textbf{D-OverDense w/o Fine-tune}: This variant drops the fine-tuning stage and tests the D-OverDense model initialized with the doc representation generated by BERT.

Observing the results in Table~\ref{tab:ablation_study}, we find that the removal of pre-training tasks will damage the results on all evaluation metrics. This indicates that the pre-training tasks indeed embed the token-level and contextual information of the documents into the model parameters. Passages tend to contain more contextual information and sampled terms focus on important tokens. In addition, discarding the fine-tuning task also impacts the two models’ performance a lot. This result confirms the effectiveness of fine-tuning with the Q-D pairs. In our DynamicRetriever, all documents are separated so that the fine-tuning task mainly captures the document-level information which is proved to be important for document retrieval and ranking.

\begin{table}[t!]
    \centering
    \caption{Exploration of Distributed Model for massive documents.}
    \begin{tabular}{p{0.3\linewidth}|p{0.15\linewidth}p{0.15\linewidth}p{0.15\linewidth}}
    \toprule
     \multirow{2}[2]{*}{Model} &  \multicolumn{3}{c}{MS MARCO} \\
        \cmidrule(lr){2-4}
        & Recall@1 & Recall@20 & MRR\\
        \midrule
        Each Group & 0.5232 & 0.8423 & 0.6445\\
        BERT & 0.1665 & 0.6321 & 0.2817\\
        Distributed Model & 0.1011 & 0.4724 & 0.1895\\
    \bottomrule
    \end{tabular}
    \label{tab:distributed_model}
\end{table}

\begin{figure}
    \centering
    \includegraphics[width=0.95\linewidth]{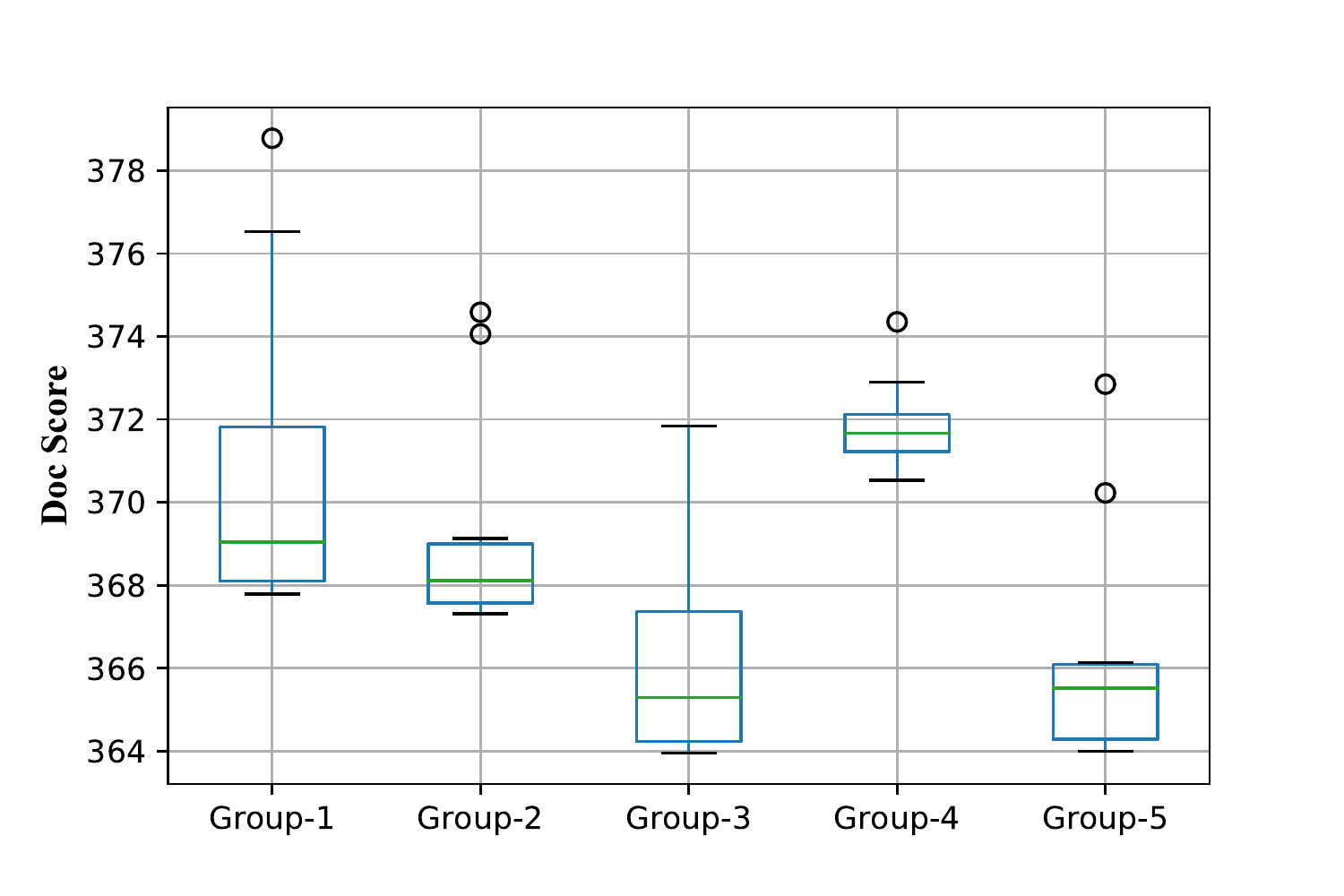}
    \caption{Score Distribution of Different Groups.}
    \label{fig:score_distribution}
\end{figure}

\subsection{Exploration of Distributed Model}
Our proposed DynamicRetriever replaces the traditional index of documents with model parameters, which naturally needs to consider the problem of how to deal with massive web documents. In Section~\ref{subsec:discussion} of this paper, we briefly discuss this challenge and provide two potential solutions. Here, we attempt to explore the method of the Distributed model. The experiments are conducted on the 3.2 million document collection of MS MARCO. We randomly divide all these documents into 32 groups where each group corresponds to 100K documents, and each group owns an individual D-OverDense model for indexing and retrieval. When testing the distributed model, given a query, the D-OverDense model of each group will retrieve and return the top 100 documents, and then these documents are merged into a ranking list according to their relevance score and returned as the final ranking result. The experimental results are shown in Table~\ref{tab:distributed_model}. We analyze the potential and existing challenges of this method as follows.

From Table~\ref{tab:distributed_model}, we can find that an individual D-OverDense model performs well on any document group, which indicates that our approach is promising to locate relevant documents accurately. However, the ranking results after merging decrease sharply and are significantly worse than the results of the classical BERT model. We analyze it is because the scale of document scores between D-OverDense models trained independently are not consistent, thus directly merging documents from various groups according to their scores would generate poor ranking results. We further compare the distribution of document scores between different groups. The comparison results illustrated in Figure~\ref{fig:score_distribution} also confirm our analysis and conjecture.

\section{conclusion}
In this paper, we propose a novel pre-training framework DynamicRetriever for document retrieval, which regards the document identifiers as tokens and train the relations from text to them. In such a framework, the semantic information of each document is stored in the model as parameters and there is no need to build an index when retrieving documents for a given query. We implement the model with two training strategies: training from scratch and training over dense vectors. Experiments on MS MARCO dataset show that our model-based IR system can improve the retrieval quality significantly, and the OverDense model demonstrates strong generalization and robustness when scaling up the corpus size. In the future, there are still many challenges, such as how to deal with the massive amount of documents at the Internet level. Therefore, model compression, multitasking and other directions are all to be explored.

\bibliographystyle{ACM-Reference-Format}
\bibliography{sample}

\end{document}